\documentclass[aip,apl,twocolumn,groupedaddress,superscriptaddress,altaffilletter,runinaddress, letter]{revtex4}
\usepackage{graphicx,epsf}
\usepackage{bm}

\begin{document}

\title{Microstrip resonator for microwaves with controllable polarization}

\author{T. P. Mayer Alegre}
\email{talegre@lnls.br} \altaffiliation{Instituto de F\'isica Gleb
Wataghin Universidade Estadual de Campinas, Campinas, SP, Brazil}

\author{A. C. Torrezan de Souza}
\altaffiliation{Massachusetts Institute of Technology, Cambridge,
Massachusetts}

\author{G. Medeiros-Ribeiro}
\email{gmedeiros@lnls.br} \affiliation{Laborat\'orio Nacional de Luz
S\'{\i}ncrotron, Caixa Postal 6192 - CEP 13084-971, Campinas, SP,
Brazil}

\date{\today}

\begin{abstract}

In this work the authors implemented a resonator based upon
microstrip cavities that permits the generation of microwaves with
arbitrary polarization. Design, simulation, and implementation of
the resonators were performed using standard printed circuit boards.
The electric field distribution was mapped using a scanning probe
cavity perturbation technique. Electron spin resonance using a
standard marker was carried out in order to verify the polarization
control from linear to circular.

\end{abstract}

\maketitle

Electron paramagnetic resonance (EPR) is a well established tool for
assessing the electron spin degree of freedom in paramagnetic
centers in a variety of systems. EPR has been traditionally used to
investigate their response to an applied magnetic field and to map
the corresponding symmetries inside crystals\cite{abragam, poole}.
More recently\cite{NV_Diamond}, optically detected magnetic
resonance (ODMR) on single centers has been used to carry out
manipulation of spin states targeting the implementation of
prototypical architectures for quantum information processing (QIP)
devices. State preparation, manipulation and read-out can be
accomplished by proper choice of microwave and optical
stimuli\cite{NV_Diamond}. Nevertheless, for adequate QIP
implementations it is important to further improve the ability to
select and tune the transitions; for example, Stark shifting levels
associated with optical transitions may allow better coupling with
optical cavities\cite{tamarat06}.

The design of microwave cavities has a direct impact on the
prospects of spin resonance apparatus applied to QIP. In addition to
improving the quality factor, which allows for greater sensitivity
\cite{poole}, polarization control of the excitation stimulus can
provide additional functionality for experimental set-ups. An
application for polarization controlled microwaves concerns the
selective excitation of spin transitions, which may impact state
preparation at zero magnetic fields\cite{alegre_07}. Also the
determination of the g-factor sign\cite{chang} requires circularly
polarized microwaves. This has been an issue of recent debate
concerning electron spins in quantum dots \cite{pryor05, bayer06},
which bears significant importance to solid state QIP. Earlier work
on generating circularly polarized microwaves employed cylindrical
cavities coupled to waveguides through an iris \cite{galt55,
clyde,suematsu,chang, filter}. Later, a square cavity pumped by two
90$^{\circ}$ out-of phase signals was successfully employed to
generate circularly polarized microwave\cite{diaz}. These
implementations required mechanical adjustments for polarization
control, which may pose difficulties for the generation of arbitrary
polarization of microwaves, or even polarization modulation.

In this work we designed and fabricated a two-port microstrip
resonator implemented by patterning a half-wavelength resonator onto
a dielectric substrate with a ground plane that allows for arbitrary
control of the microwave polarization. This was experimentally
verified by carrying out EPR experiments with samples containing a
conventional marker with known g-factor.

\begin{figure}[ht!]
\centerline{\epsffile{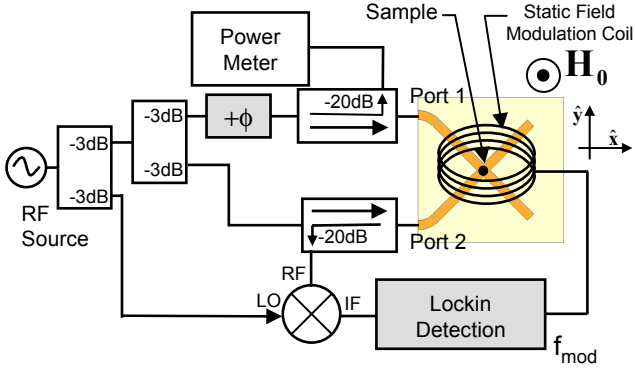}}%
\caption{Experimental setup for EPR measurement. The cavity was
driven with a phase difference $\phi$ between each port, set by a
phase shifter with negligible insertion loss. The reflected signal
from port 2 is mixed with the local oscillator and detected at the
modulation frequency of the static magnetic field ($\rm{
f_{mod}}\approx 5{\rm~kHz}$) by a Lock-in amplifier. A power meter
is used to monitor the input power.} \label{setup}
\end{figure}

Resonators were fabricated after simulation utilizing Ansoft HFSS
software \cite{ansoft} on printed circuit boards (PCB) through
conventional photolithography. PCBs (Rogers model RO 3203) with a
relative permittivity $\varepsilon_r=3$ were utilized. The spatial
distribution of the z component of the electric field, $E_z$, was
mapped by means of a scanning probe cavity perturbation technique. A
300$\mu$m diameter, 8 mm long electrochemically sharpened tungsten
tip was scanned 250$\mu$m over the resonator surface and the cavity
response was concomitantly monitored with a vector network analyzer.
The intensity of the field can be directly related to the shift of
the natural frequency of the cavity when perturbed by the metallic
tip\cite{rf_book,maier}. The EPR set-up consisted of a standard dc
magnet with field modulation supplied by an additional coil, and a
solid state microwave source. The two ports of the resonator were
pumped independently and the phase $\phi$ between the excitation
signals could be adjusted by a phase shifter with negligible
insertion loss. The reflected signal was Lock-in detected using a
mixer operating as a linear detector\cite{poole} (Fig.~\ref{setup}).
The experiments were performed at room temperature utilizing DPPH
(1,1-diphenyl-2-picrylhydrazyl, g=2.0036$\pm$0.0003\cite{poole}) as
the spin marker.

\begin{figure}[ht]
\centerline{\epsffile{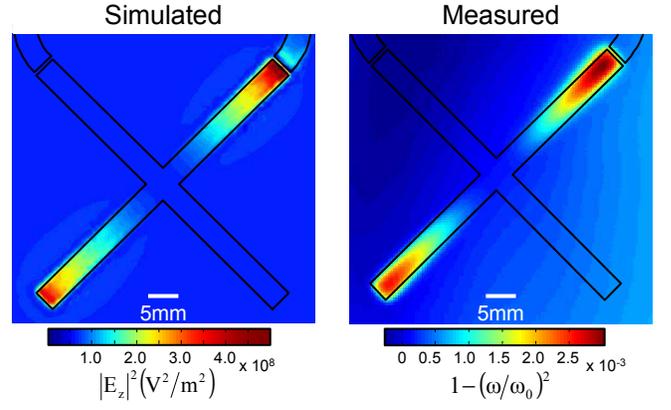}}%
\caption{ Calculated and measured (displayed as
$1-(\omega/\omega_0)^2$) distribution of the electric field parallel
to z-axis. The measured electric field was done through a thin
needle (with length of ~8mm and radius of ~300$\mu$m) placed
parallel to z-direction. Each point in the measured field map is a
400x400$\mu$m$^2$ square.} \label{field}
\end{figure}

By placing two half-wavelength resonators orthogonally, one can
control the direction and polarization of the fields at their
intersection by adjusting the amplitude and phase of the excitation
signal at each port. The zeroth order mode of the proposed resonator
consists of two degenerate and orthogonal modes ({\it i.e.,} the
superposition of the quasi ${\bf TEM_{01}}$ and ${\bf TEM_{10}}$
modes) that can be pumped independently by two oscillators with an
arbitrary phase difference, thereby creating the desired degree of
polarization. For the half-wavelength resonator a first estimate for
the resonance frequency $f$ can be evaluated by $f\approx nc/(2 l
\sqrt{\varepsilon_r})$, where $\rm n$ is the cavity mode, $\rm c$
the speed of light in vacuum\cite{johansson74, silesbee91}. In the
case of the two port resonator, the first mode frequency should not
be too far from this estimate. For a resonator fabricated on the PCB
with an arm length of 5.5 cm we obtained the first mode resonance at
a frequency of $\rm 1.71GHz$. The electric field $E$ can be
associated to the shift in the resonance frequency of the cavity by
$1-(\omega/\omega_0)^2=\beta E^2$, where $\omega$, $\omega_0$ and
$\beta$ are the perturbed frequency, natural frequency and a
geometrical factor that takes into account the shape of the
perturbing object and the volume occupied by the electromagnetic
field\cite{rf_book,maier}. Fig.~\ref{field} shows the spatial
distribution of the electric field $E_z$ for one of the degenerate
modes, calculated and imaged at 1.71GHz (displayed as
$1-(\omega/\omega_0)^2$). This mode has two important
characteristics: a) it has a node for the electric field at the
center of the cavity, which is essential for EPR experiments in
quantum dots \cite{klaus06,koppens06}; this fact also enables the
investigation of highly conductive samples without loading the
cavity; b) there is a good isolation (50dB) between this mode and
the orthogonal, degenerate mode (not shown); this permits
independent pumping of the resonator, which is crucial for
polarization control. The agreement between the experiment and
calculation is qualitative differing by a multiplying factor $\beta$
\cite{rf_book, maier}; nevertheless, the scanning probe perturbation
method allows us to identify the ideal spot to place the sample.

\begin{figure}[t!]
\centerline{\epsffile{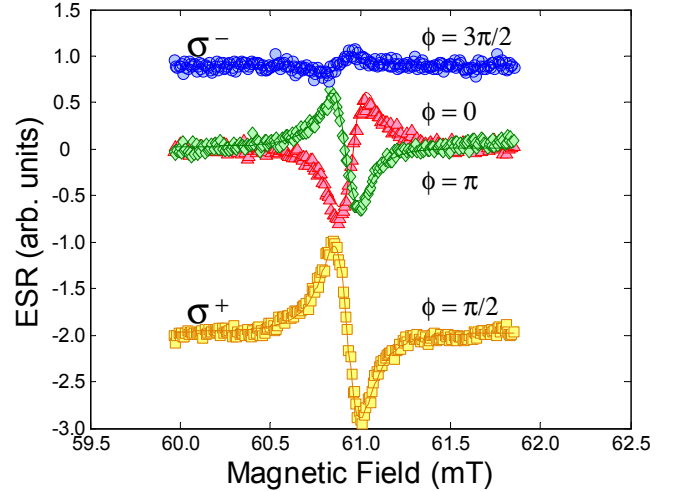}}%
\caption{ EPR measurements for $\phi=0$(red solid triangles),
$\phi=\pi/2$(yellow solid squares) , $\phi=\pi$ (green filled
diamond) and $\phi=3\pi/2$ (blue solid circles).  The $\sigma^{+}$
and $\sigma^{-}$ polarization take place at $\phi=\pi/2$ and
$\phi=3\pi/2$ respectively. For the other two phases one has linear
polarization parallel to the $x$ and $y$ axis, respectivelly (see
Fig.~\ref{setup}). The solid lines are fits to the resonances,
represented by the derivative of a lorenztian lineshape.}
\label{esr}
\end{figure}

In order to evaluate the degree of polarization, we performed EPR
experiments. By measuring the EPR signal as a function of the phase
difference $\phi$ between the microwave stimuli at the two ports,
one can determine whether the polarization is linear, circular or
elliptical. The DPPH sample was positioned at the center of the
cavity where the circularly polarized microwave reaches the maximum
degree of polarization and intensity. Fig.~\ref{esr} shows the EPR
signal for DPPH for $\phi = 0, \pi, 3\pi/2$ and $\pi/2$. For $\phi =
0, \pi$, the signal amplitude does not change, which confirms linear
polarization for the microwaves; the orientation of ${\bf B_1}$ is
parallel and perpendicular to the $x$-axis (see Fig.~\ref{setup}),
respectively. This creates a $180^\circ$ phase shift in the detected
signal. For $\phi =\pi/2$, the amplitude of the EPR signal is
maximum, in accord with $\sigma^+$ polarization, since the g-factor
for DPPH is positive. As expected, for $\phi =3\pi/2$ ($\sigma^-$
polarization), the EPR signal is minimum.

One can define the degree of polarization by
$P_{\sigma^+}=|A_{\sigma^+}-A_{\sigma^-}|/(A_{\sigma^+}+A_{\sigma^-})$,
where $A_{\sigma^{\pm}}$ are the signal amplitudes for
$\sigma^{\pm}$ polarization. We find $P_{\sigma^+}\cong 80\%$. The
fact that the microwaves are not $100\%$ polarized can be attributed
to: a) slight differences of power and phase at each port due to
imperfections of the connectors and cables; b) the geometry of the
intersection of the two arms is not circular, but rather a square;
this will produce spatial harmonics and the circular polarization is
present only at one point at the very center of the resonator; c)
imperfections in the fabrication of the resonator, in particular the
coupling gap (200$\mu$m in width), which can produce phase
differences. These problems can be eliminated by external trimming
of the amplitude and phase of the incoming microwaves and using
smaller samples.

\begin{figure}[t!]
\centerline{\epsffile{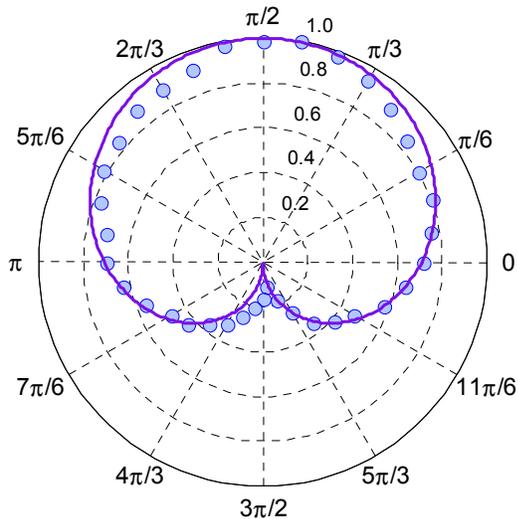}}%
\caption{ Polar plot showing the EPR signal as a function of the
phase difference between the two ports. For a phase difference of
$\pi/2$ the resulting microwave field polarization is $\sigma^+$
which produces a maximum for the EPR signal. On the other hand for a
phase difference of $3\pi/2$ one has a $\sigma^-$ polarization,
generating a negligible EPR signal.} \label{polar}
\end{figure}

Finally, Fig.~\ref{polar} shows the EPR signal amplitude $A(\phi)$
as a function of the phase difference $\phi$ between the microwave
stimuli at each arm. This dependence can be shown to be
$A(\phi)/A_{\sigma^+}= \sqrt{(1+sin(\phi))/2}$, which is represented
in the polar plot by the solid line. This graph demonstrates the
ability of creating arbitrary polarization, from linear to circular,
with a degree of polarization better than 80\% for any desired
combination of linear and circular polarization.

In summary, we successfully designed and implemented a resonator
that allows one to produce arbitrarily polarized microwaves. This
resonator can in principle produce microwaves with time-dependent
polarization modulation, by either adjusting the phase or the
amplitudes at each arm. We also employed an electric field mapping
technique to spatially locate the resonator modes, thus allowing an
optimum choice of a suitable mode for polarization control. For the
modes where the electric field has a node at the cavity mid-point,
the loading of the resonator is minimal, which can be quite
advantageous in applications where an electric contact to the sample
is required. Additionally, by choosing a material with high
dielectric permittivity, the resonators can be made very compact
which can be easily integrated in systems where space is a
constraint. These devices represent interesting perspectives for the
implementation of experiments which require polarization control of
microwaves, including for example the determination of the
electronic g-factor sign and transition selection rules.

We acknowledge the financial support by FAPESP process number
04/01286, CNPq and HP-Brazil. The authors would like to thanks the
technical assistance of M. H. Piazzetta and A. L. Gobbi from the
micro-fabrication facility at LNLS for help with the fabrication of
the resonators.

\end{document}